# Emergence of high-mobility carriers in topological kagome bad metal $Mn_3Sn$ by intense photoexcitation

Takuya Matsuda[1,*,†], Tomoya Higo[1,2], Kenta Kuroda[1,3,4], Takashi Koretsune[5], Natsuki Kanda[1,‡], Yoshua Hirai[2], Hanyi Peng[2], Takumi Matsuo[2], Cedric Bareille[1], Andrey Varykhalov[6], Naotaka Yoshikawa[2], Jun Yoshinobu[1], Takeshi Kondo[1], Ryo Shimano[2,7], Satoru Nakatsuji[1,2,8,9] and Ryusuke Matsunaga[1,8,*]

[1]*The Institute for Solid State Physics, The University of Tokyo, Kashiwa, Chiba 277-8581, Japan.*

[2]*Department of Physics, The University of Tokyo, Bunkyo-ku, Tokyo, 113-0033, Japan.*

[3]*Graduate School of Advanced Science and Engineering, Hiroshima University, 1-3-1 Kagamiyama, Higashi-Hiroshima 739-8526, Japan.*

[4]*International Institute for Sustainability with Knotted Chiral Meta Matter (WPI-SKCM2), Hiroshima University, Higashi-hiroshima, Hiroshima 739-8526, Japan.*

[5]*Department of Physics, Tohoku University, Sendai, 980-8577, Japan.*

[6]*Helmholtz-Zentrum Berlin für Materialien und Energie, Elektronenspeicherring BESSY II, Albert-Einstein-Strasse 15, 12489 Berlin, Germany.*

[7]*Cryogenic Research Center, The University of Tokyo, Bunkyo-ku, Tokyo, 113-0032, Japan.*

[8]*Trans-scale Quantum Science Institute, The University of Tokyo, Bunkyo-ku, Tokyo, 113-0033, Japan.*

[9]*Institute for Quantum Matter and Department of Physics and Astronomy, Johns Hopkins University, Baltimore, 21218 Maryland, USA.*

*e-mail: matsuda@mp.es.osaka-u.ac.jp, matsunaga@issp.u-tokyo.ac.jp

†Present address: Department of Materials Engineering Science, Osaka University, 1-3 Machikaneyama, Toyonaka, Osaka 560-8531, Japan.

‡Present address: Present address: RIKEN Center for Advanced Photonics, RIKEN, 2-1 Hirosawa, Wako, Saitama 351-0198, Japan.

## Abstract

Kagome-lattice materials offer novel playgrounds of exploring topologically nontrivial states of electrons under influence of many-body interactions. A noncollinear kagome antiferromagnet $Mn_3Sn$ has attracted particular interest for application in spintronics owing to the large anomalous Hall effect related to the Weyl dispersion near the Fermi energy. In addition, strong electronic correlation suggesting the Kondo physics has also been implied. However, the effect of correlation on the band topology and their interplay remains elusive. Here, we investigate nonequilibrium Hall transport in a photoexcited $Mn_3Sn$ using time-resolved terahertz Faraday rotation spectroscopy. In equilibrium, $Mn_3Sn$ is a bad metal close to the Mott-Ioffe-Regal limit with low carrier mobility, and thus only the anomalous Hall effect is discerned. By contrast, intense photoexcitation beyond an approximate threshold gives rise to a clear cyclotron resonance, namely the normal Hall effect, indicating the emergence of unusual carriers with 50 times lighter effective mass and 40 times less scattering. The lifetime of high-mobility carriers as long as a few tens of picoseconds and a threshold-like behavior for the pump fluence are hardly explained by contribution of photoexcited hot carriers. Instead, the emergence of unusual carriers may be accounted for by dielectric screening of the on-site Coulomb interaction by high-density delocalized photocarriers. A possible role of electronic correlation in equilibrium transport in $Mn_3Sn$ beyond the single-particle picture is discussed.

## I. INTRODUCTION

Many-body interactions between electrons induce diverse phases of matter, which has been a central research subject in modern condensed matter physics [1]. In addition, a topological perspective of the band structure has ignited a paradigm shift, even within the single-electron approximation, unveiling nontrivial properties of relativistic Dirac or Weyl fermions [2]. Accordingly, the effect of electron correlation on topological materials has been studied to understand the interplay between many-body physics and topology, leading to emergent phases and collective phenomena [3–6], including topological superconductivity [7] or new ordered phases [8–10]. However, topologically nontrivial phases under strong electron correlations are elusive in experiments.

Kagome lattice often serves as a platform for topological physics under electronic correlations [11]. Particularly, a noncollinear kagome antiferromagnet $Mn_3Sn$ shows a large anomalous Hall effect (AHE) at room temperature [12] owing to the ferroic order of cluster magnetic octupole as shown in Fig. 1(a) [13,14], attracting significant attention for application in spintronics [15,16]. The band structure calculation using the density functional theory (DFT) calculations in Fig. 1(b) has expected the presence of Weyl nodes near the Fermi energy [17–20], which was experimentally confirmed by chiral anomaly [21]. However, a clear spectroscopic signature of Weyl cone dispersions has evaded observation in angle-resolved photoemission spectroscopy (ARPES) due to strong correlation between the Mn 3*d* electrons [21]. Transport experiment has also reported a short mean-free-path close to the lattice constant [22], suggesting that the system is near the Mott-Ioffe-Regal limit [23]. In addition, an insulating behavior with substitution of Sn with Mn atoms [24] and a sharp Fano-shaped spectrum observed in scanning tunneling spectroscopy [25] have also suggested that the system is close to a correlation-induced gapped phase. Despite these smoking gun experiments, however, the effect of strong correlation on their nontrivial band topology remains unresolved, and most studies have still treated this material within the single-electron approximation.

An experimental observable which reflects both the topological nature and correlation of electrons may be the off-diagonal Hall conductivity $\sigma_{xy}$. The intrinsic AHE in magnets is driven by the Berry curvature [26], which behaves as a pseudo magnetic field in momentum space and is enhanced around the Weyl nodes. In addition, when carriers are driven by the Lorentz force under a magnetic field in real space, the normal Hall effect (NHE) occurs. The NHE is sensitive to carrier mobility, which is significantly suppressed under strong electronic correlation. Thus, in-depth exploration of the AHE and NHE with changing a physical parameter in a single material would unveil the effect of many-body physics on the band topology. For this purpose, in contrast to the conventional Hall measurement in dc transport, the Faraday rotation spectroscopy using terahertz (THz) pulse would be highly beneficial because it provides the spectral profile of Hall conductivity, $\tilde{\sigma}_{xy}(\omega)$. The NHE gives rise to a dissipative peak in Im $\tilde{\sigma}_{xy}(\omega)$ at the cyclotron resonance frequency $\omega_c/2\pi$ depending on an external magnetic field [27–29] as shown in Fig. 1(c). By contrast, the $\tilde{\sigma}_{xy}(\omega)$ dominated by the intrinsic AHE tends to be non-dissipative in the THz frequency with a nearly flat spectral shape at room temperature [30,31] as shown in Fig. 1(d). Thus, the spectral information helps to distinguish the microscopic origin of Hall current [32,33].

Furthermore, the sub-picosecond time resolution for detecting a THz pulse allows us to track ultrafast change in $\tilde{\sigma}_{xy}(\omega)$ away from equilibrium [34,35]. When high-density photocarriers are excited by intense laser pulses, it has been theoretically expected that the Coulomb interaction can be screened by delocalized photocarriers [36–41], which may trigger the emergence of hidden states of matter as anticipated to stabilize a hidden magnetic Weyl semimetal phase in pyrochlore iridates [39]. The effect of photocarrier screening on the Coulomb interaction has been intensively investigated for electron-hole systems in semiconductors [42–44], Mott insulators [45–47] and an excitonic insulator [48,49]. Recently, even in metallic systems, the photocarrier screening for Lifshitz transition of a correlated Weyl semimetal [50] and band renormalization of a cuprate superconductor [51] has also been demonstrated. Compared to Floquet engineering where electron states are coherently coupled with light field [52], the photocarrier screening is beneficial to explore a nonequilibrium phase because it can sustain even after turning off the light [39]. Therefore, probing the transient change in $\tilde{\sigma}_{xy}(\omega)$ after intense photoexcitation in correlated topological materials using THz pulses will provide fruitful information for exploring dynamical control of electronic correlation.

In this work, using time-resolved THz Faraday spectroscopy, we investigate nonequilibrium carrier transport in a photoexcited $Mn_3Sn$ thin film under a magnetic field. A previous magneto-THz spectroscopy for $Mn_3Sn$ in equilibrium has reported that any signature of the NHE is not identified up to 7 T [53], which has been attributed to a suppressed carrier mobility by electronic correlation. Our work reveals that an intense near-infrared photoexcitation beyond a threshold induces a clear NHE in the form of cyclotron resonance, exhibiting an emergence of unusual carriers with a higher mobility than in equilibrium by three orders of magnitude. Concomitantly, a signature of the less scattered carriers was also discerned in the longitudinal conductivity spectrum after intense photoexcitation. A possible origin of the emergence of unusual carriers is discussed.

This paper is organized as follows. First, we introduce the methods in Sec. II. After the information of sample characterization (Sec. II A) and DFT calculation (Sec. II B), we describe the experimental setup of optical pump–THz Faraday probe spectroscopy under static magnetic fields (Sec. II C) to investigate transient Hall conductivity spectrum $\tilde{\sigma}_{xy}(\omega)$ in a $Mn_3Sn$ thin film. We also show the experimental setup of optical pump–THz probe spectroscopy in reflection geometry without a magnetic field (Sec. II D) to investigate transient longitudinal conductivity spectrum $\tilde{\sigma}_{xx}(\omega)$ with a better signal-to-noise ratio. In Sec. III A, we show the experimental results of $\tilde{\sigma}_{xy}(\omega)$ after intense photoexcitation, indicating the emergence of unusual carriers with less scattering and a much smaller effective mass compared with those in equilibrium. In Sec. III B, we confirm that the unusual carriers also appear in $\tilde{\sigma}_{xx}(\omega)$ after intense photoexcitation. In Sec. III C, we show the dynamics of $\tilde{\sigma}_{xy}(\omega)$ and $\tilde{\sigma}_{xx}(\omega)$ where the lifetime of unusual carriers is as long as a few tens of picoseconds, much longer than an expected relaxation of photoexcited hot electrons. A possible origin of unusual, less scattered carriers is discussed from the viewpoint of dielectric screening of the Coulomb interaction by photoinjected delocalized carriers.

## II. METHODS

### A. Sample characterization

For sample fabrication, we deposited the polycrystalline $Mn_3Sn$ film (20 nm) [54] and $AlO_x$ (5 nm) passivation layer on quartz ($SiO_2$) substrates by dc and rf magnetron sputtering from $Mn_{2.7}Sn$ and $Al_2O_3$ targets in a chamber with a base pressure of $< 5 \times 10^{-7}$ Pa, respectively. The stack of $SiO_2/Mn_3Sn/AlO_x$ was fabricated at room temperature, and it was subsequently annealed at 600 °C for 30 minutes. The sputtering power and Ar gas pressure were 60 W and 0.9 Pa for $Mn_3Sn$, and 100 W and 0.2 Pa for $AlO_x$. The composition of the $Mn_3Sn$ layer was determined to be $Mn_{3.06(1)}Sn_{0.94(1)}$ by scanning electron microscopy-energy dispersive X-ray spectrometry. The temperature dependence of the remnant Hall resistivity $\rho_{xy}(B = 0\ \mathrm{T})$ in dc transport for the same sample was reported in a previous paper [34]. Below the Néel temperature at $T_N$~430 K, $\rho_{xy}$ grows because of broken time-reversal symmetry in the noncollinear antiferromagnetic phase with the cluster magnetic octupole. Below 220 K, $\rho_{xy}$ in turn starts to decrease due to the transition into the helical phase [55]. For our pump–probe spectroscopy, we set the base temperature of 220 or 300 K. At both temperatures, the $Mn_3Sn$ film is in the cluster octupole phase in equilibrium. The important difference between 220 and 300 K is whether the lattice temperature is expected to exceed $T_N$ after intense photoexcitation, as discussed later.

### B. Band structure mapping

Figure 1(b) shows the electronic band structure in $Mn_3Sn$ obtained by DFT calculation. The DFT calculations were performed within the generalized gradient approximation (GGA) [56] using the quantum-ESPRESSO package [57]. The pseudopotentials with the projector augmented wave scheme [58] were used. For discussion, we also use the band dispersion data obtained by ARPES for a bulk single crystal $Mn_3Sn$ reported in a previous paper [21]. The ARPES experiment was performed at $1^2$ beamline of BESSY II. The photoelectrons were acquired by hemispherical analysers R8000 (ScientaOmicron). The overall energy resolution was set to better than 30 meV, and the angular resolution was in 0.3 deg. The bulk $Mn_3Sn$ crystals were cleaved at approximately 60 K, exposing flat surfaces corresponding to the (0001) cleavage plane. The sample temperature was kept at 60 K during the measurement.

### C. Time-resolved Hall conductivity spectrum measurement

Figure 1(e) shows a schematic of the optical pump–THz Faraday probe spectroscopy setup with a superconducting magnet. The magnetic field was normal to the sample surface to orient the magnetization direction. All the pump–probe experiments for the Hall conductivity spectrum were performed under a static magnetic field because the Hall conductivity is closely related to the magnetic structure, which is largely perturbed by a strong pump pulse due to laser heating. In the absence of a magnetic field, the magnetic domain can be randomly oriented in the cooling process. To reset the magnetic structure for arrivals of each pump pulse, a static magnetic field $|B|$ > 1 T is required. The sample temperature was set to be 220 K, where the Weyl nodes near the Fermi energy are fully developed, leading to the Berry curvature-induced AHE. This temperature was also low enough to stabilize the sample temperature for the operation of the superconducting magnet.

An output of a regenerative amplifier with the 1.55-eV photon energy, the 40-fs pulse duration, and the 1-kHz repetition rate was divided into three parts for the optical pump pulse, generation of the THz probe pulse, and the sampling pulse. The THz pulse was generated by optical rectification in a large-aperture 0.3 mm-thick-GaP crystal. The THz pulse was linearly polarized along the $x$-direction by using a wire-grid polarizer (WGP1). The sample was covered by a tapered holder with 6-mm diameter and mounted in the superconducting magnet with pairs of diamond inner windows and $SiO_2$ outer windows. The incident THz pulse was focused onto the sample by a parabolic mirror with an effective focal length of 190 mm. After going through two other polarizers (WGP2 and WGP3), the THz field was detected with the sampling pulse by electro-optic sampling in a 2-mm-thick ZnTe crystal. By scanning the probe delay $t_{\mathrm{probe}}$ to the sampling pulses for the THz probe, the THz pulse waveform was obtained and transformed to the frequency domain for spectral analysis.

The optical pump pulse irradiated the sample with a controllable pump delay $t_{\mathrm{pump}}$ to the sampling pulse. The optical pump beam had a Gaussian profile with $1/e^2$ diameter of 6 mm, which ensured spatial uniformity of fluence on the probed region. For the optical pump and polarization-resolved THz spectroscopy measurement, the angle of WGP3 was fixed to block the $x$-component THz field, and the ZnTe crystal was set to maximize the detection efficiency of the $y$-component field. The angle of the WGP2 was changeable during the measurement. Further details in the analysis are represented in Ref. [34].

The longitudinal conductivity spectrum $\tilde{\sigma}_{xx,\mathrm{w/}}(\omega)$ with the pump is expressed as

$$\tilde{\sigma}_{xx,\mathrm{w/}}(\omega) = \left[\tilde{E}_{x,\mathrm{w/o}}(\omega)/\tilde{E}_{x,\mathrm{w/}}(\omega)\left(1+n_s+\tilde{\sigma}_{xx,\mathrm{w/o}}(\omega)Z_0 d\right)-(1+n_s)\right]/(Z_0 d),$$

where $Z_0$=377 Ω is the vacuum impedance, $d$ is the film thickness, $n_s$ is the refractive index of the substrate, and $\tilde{\sigma}_{xx,\mathrm{w/o}}(\omega)$ is the longitudinal conductivity spectra without the pump. The Hall conductivity spectrum $\tilde{\sigma}_{xy,\mathrm{w/}}(\omega)$ with the pump is also obtained from the relation,

$$\tilde{\sigma}_{xy,\mathrm{w/}}(\omega) = \tilde{\theta}_{F,\mathrm{w/}}(\omega)\left(1+n_s+\tilde{\sigma}_{xx,\mathrm{w/}}(\omega)Z_0 d\right)/(Z_0 d),$$

where $\tilde{\theta}_{F,\mathrm{w/}}(\omega)$ are the complex Faraday rotation spectra with the pump. We applied magnetic fields of $+B$ and $-B$, and obtain $E_x$ and $E_y$ of the THz field as even and odd components for $B$, respectively, to evaluate $\tilde{\sigma}_{xy}(\omega)$.

### D. Time-resolved longitudinal conductivity spectrum measurement

We also built up another optical pump–THz probe spectroscopy setup in reflection geometry in free space at room temperature. This system was used to observe $\sigma_{xx}(\omega)$ signal with a better signal-to-noise ratio and stronger pump fluences as discussed later. For this experiment, another regenerative amplifier system with the pulse energy of 7 mJ, the pulse duration of 35 fs, the repetition rate of 1 kHz, and the photon energy of 1.55 eV were used. The output of the laser was divided into three parts: each for the optical pump, the generation of the THz probe, and the sampling pulse for the THz probe, respectively. The optical pump beam had a Gaussian profile with $1/e^2$ diameter of 8 mm. The probe THz pulse was generated by optical rectification in a large aperture 0.3-mm-thick GaP crystal and the reflected THz pulse was detected by the electro-optic sampling in another 0.3-mm-thick GaP crystal with the sampling pulse. For the longitudinal conductivity measurement, applying a static magnetic field was not necessarily because the response of the unusual carriers in the longitudinal conductivity appears in the same way even when the magnetic domains were randomized.

## III. RESULTS

### A. Time-resolved Hall conductivity spectra

Figures 2(a) and 2(b) show the real- and imaginary-part $\tilde{\sigma}_{xy}(\omega)$, respectively, at 220 K for various pump fluences $I_p$ at a fixed pump delay of $t_{\mathrm{pump}}$ = 1.2 ps under $B$ = 2 T. The black curves show $\tilde{\sigma}_{xy}(\omega)$ in equilibrium. Re $\tilde{\sigma}_{xy}(\omega)$, corresponding to a non-dissipative current, shows a flat spectrum with the value as large as 20 $\Omega^{-1}$ $cm^{-1}$. By contrast, the dissipative response Im $\tilde{\sigma}_{xy}(\omega)$ is negligibly small. This feature is consistent with the AHE in $Mn_3Sn$ previously observed [30]. It is known that the spectral shape of $\tilde{\sigma}_{xy}(\omega)$ with the intrinsic origin reflects the electron band structure [32]. But the small photon energy in the THz frequency and the large thermal energy at room temperature makes the $\omega$ dependence less noticeable and hold $\tilde{\sigma}_{xy}$ in the dc limit [30, 31]. For a weak pump fluence $I_p$ = 0.3 mJ $cm^{-2}$, $\tilde{\sigma}_{xy}(\omega)$ hold its spectral shape and only the real part decreases. A recent study revealed that this ultrafast suppression of the AHE is ascribed to the change in the electron temperature and is well explained by the intrinsic Berry-curvature mechanism [34].

The present work focuses on the results of much stronger pump fluences $I_p$ = 1.4 and 2.2 mJ $cm^{-2}$ (green and orange), under which Re $\tilde{\sigma}_{xy}(\omega)$ drastically changed, and the low-frequency part flipped its sign in Fig. 2(a). Concomitantly, a peak structure appeared in Im $\tilde{\sigma}_{xy}(\omega)$ in Fig. 2(b). As $I_p$ increases, these features were further enhanced. We also investigated $B$ dependence of $\tilde{\sigma}_{xy}(\omega)$ under the strong pump fluence. Figures 2(c) and 2(d) show the real- and imaginary-part $\tilde{\sigma}_{xy}(\omega)$, respectively, at 220 K with $I_p$ = 2.2 mJ $cm^{-2}$ at $t_{\mathrm{pump}}$ = 1.2 ps. As $B$ increases, the spectral feature was notably pronounced and the peak in Im $\tilde{\sigma}_{xy}(\omega)$ shifted to the higher-frequency side. The blue shift is more clearly seen in Re $\tilde{\sigma}_{xy}(\omega)$ in Fig. 2(c), where the zero-crossing point shifted to the higher-frequency side as $B$ increases. The results suggest that $\tilde{\sigma}_{xy}(\omega)$ after the high-density photoexcitation is no longer dominated by the AHE.

Considering the NHE, *i.e.*, the cyclotron resonance, we fitted the experimental data by using the following equation,

$$\tilde{\sigma}_{xy}(\omega) = \frac{\varepsilon_0 \omega_p^2 \omega_c}{(\omega + \mathrm{i}/\tau_c)^2 - \omega_c^2} + \sigma_{\mathrm{AHE}}, \tag{1}$$

where $\varepsilon_0$ is the vacuum permittivity, $\omega_p/2\pi$ is the plasma frequency, $\omega_c/2\pi$ is the cyclotron frequency, and $\tau_c$ is scattering time in the cyclotron resonance. $\sigma_{\mathrm{AHE}}$ is the $\omega$-independent anomalous Hall conductivity (See Appendix A). Figure 2(e) shows the obtained parameters $\omega_c/2\pi$ and $\tau_c$ as a function of $B$. The $\omega_c/2\pi$ showed a linear dependence on $B$, which is fully consistent with the cyclotron resonance relation, $\omega_c = eB/m^*$, where $e$ is the elementary charge. From this result, the effective mass $m^*$ was estimated to be $\sim 0.2\, m_e$ ($m_e$ the mass of bare electrons). We also evaluated the density of the carriers contributing to the cyclotron resonance from the relation $N_c = \varepsilon_0 m^* \omega_p^2 / e^2$. Figure 2(f) shows $N_c$ obtained from the fitting as a function of $I_p$, showing a threshold-like behavior at $I_p$~1 mJ cm$^{-2}$.

Compared with the equilibrium transport in $Mn_3Sn$, the carriers contributing to the cyclotron resonance are quite unusual; According to the dc Hall coefficient, the original carrier density is $2\times10^{22}$ cm$^{-3}$ [12,22]. Considering the dc conductivity of ~4,000 $\Omega^{-1}$cm$^{-1}$, the original carrier mobility is as low as 1 cm$^2$V$^{-1}$s$^{-1}$. The scattering time of ~8 fs observed by THz spectroscopy [30,53] gives that $m^*$ should be as large as 11 $m_e$. By contrast, the unusual carriers that emerge far away from equilibrium in this work show $m^* {\sim} 0.2\, m_e$ and $\tau_c \sim 300$ fs, which are respectively 50-times smaller and 40-times longer than those in equilibrium, resulting in much higher mobility of $\mu \sim 2{,}600$ cm$^2$V$^{-1}$s$^{-1}$, enhanced by a factor of 2,000. Note that the density of unusual carriers in Fig. 2(f) is five orders of magnitude smaller than that of the original carriers in equilibrium. Nevertheless, the high-mobility carriers can yield a much larger NHE because the normal Hall conductivity in the dc limit is proportional to $\mu^2$ (See Appendix A).

### B. Time-resolved longitudinal conductivity spectra

We further examined whether the unusual carriers appear in the longitudinal transport under the high-density photoexcitation. The longitudinal conductivity spectra $\tilde{\sigma}_{xx}(\omega)$ under the strong pump is expected to show two components from (i) the original carriers in equilibrium and (ii) the unusual high-mobility carriers. Thus, we assume two Drude oscillators:

$$\tilde{\sigma}_{xx}(\omega) = -\varepsilon_0 \left( \frac{\omega_{p,1}^2}{\mathrm{i}\omega - 1/\tau_1} + \frac{N_c e^2}{m^* \varepsilon_0} \frac{1}{\mathrm{i}\omega - 1/\tau_2} \right). \quad (2)$$

The first term shows the original carriers with the plasma frequency of $\omega_{p,1}/2\pi$ = 350 THz and the scattering time of $\tau_1$ = 8 fs [30,53]. The second term shows the contribution of unusual carriers, and we set $N_c$ = $10^{17}$ cm$^{-3}$, $m^*$ = 0.2 $m_e$, and $\tau_2$ = 300 fs from the results in Fig. 2. Figure 3(a) shows the result of the model calculation. The thin black curves show only the first term in Eq. (2). The thick red and blue curves show the real and imaginary parts including both two terms. Because of the long scattering time $\tau_2$, the contribution of the unusual carriers in $\mathrm{Re}\, \tilde{\sigma}_{xx}(\omega)$ could appear only slightly below 2 meV on a large offset of the original carriers, which is out of our probe frequency window. Due to the Kramers–Kronig relation, however, the long-$\tau$ Drude response is accompanied by an increase in the imaginary part, as shown by the blue curve in Fig. 3(a), which is considerably broader in frequency than the real part and thus may be observed in our THz spectroscopy.

To investigate the longitudinal conductivity, we conducted a pump–probe spectroscopy at 220 and 300 K with a better signal-to-noise ratio. Figure 3(b) shows $\tilde{\sigma}_{xx}(\omega)$ at 300 K at 0.64 ps immediately after pump irradiation with a fluence of 3.3 mJ cm$^{-2}$. We observed that $\mathrm{Im}\,\tilde{\sigma}_{xx}(\omega)$ increases toward a lower frequency both at 220 and 300 K, substantiating the appearance of long-$\tau$ carriers in the Drude response. Figure 3(c) shows the pump-induced change, $\mathrm{Im}\,\Delta\sigma_{xx}(\omega)$, at $t_{\mathrm{pump}}$ = 0.64 ps with various $I_p$. Figure 3(d) shows the pump fluence dependence of the spectrally integrated $\mathrm{Im}\,\Delta\sigma_{xx}$ from 2.4 to 6.1 meV at $t_{\mathrm{pump}}$ = 0.64 ps. The increase in $\mathrm{Im}\,\Delta\sigma_{xx}$ indicates a threshold behavior at $I_p$~1 mJ cm$^{-2}$, which is reasonably consistent with the result of the unusual carrier density $N_c$ in Fig. 2(f). Thus, the emergence of unusual high-mobility carriers was confirmed in both the Hall and longitudinal responses.

### C. Dynamics of longitudinal and Hall conductivity spectra

We also investigated the temporal evolution of the unusual carriers. Figure 4(a) shows the Hall conductivity spectra $\tilde{\sigma}_{xy}(\omega)$ at $t_{\mathrm{pump}}$ = 1.2 and 50 ps, respectively, at the base temperature of 220 K with $I_p$ = 1.4 mJ cm$^{-2}$. Immediately after the pump at $t_{\mathrm{pump}}$ = 1.2 ps, a pump-induced cyclotron resonance emerges. Intriguingly, even at $t_{\mathrm{pump}}$ = 50 ps after the pump, we observed that the feature of the cyclotron resonance noticeably survives. This long lifetime was also confirmed in the dynamics of $\tilde{\sigma}_{xx}(\omega)$ as shown in Fig. 4(b). The results indicate that the unusual carriers have a long lifetime over 10 ps after the pump. By contrast, Fig. 4(c) shows the dynamics of $\mathrm{Im}\,\Delta\sigma_{xx}(\omega)$ at the base temperature of 300 K with 3.3 mJ cm$^{-2}$, which indicates that the feature of unusual carriers rapidly vanishes within a few picoseconds.

## IV. DISCUSSION

Here we discuss the origin of unusual, light-mass and less-scattered carriers. If the photoexcited hot carriers could remain at high-energy bands with a light effective mass, they might contribute to cyclotron resonance. However, the hot carriers must lose their kinetic energies rapidly due to the scattering as fast as several femtoseconds [34], which is clearly different from the unusual carriers observed even 50 ps after the photoexcitation in this work as shown in Figs. 4(a) and 4(b). In addition, the threshold-like behavior with respect to the pump fluence in Figs. 2(f) and 3(d), and the scattering time as long as 300 fs in Fig. 2(e), are also difficult to be explained by a simple redistribution of electron into higher-energy bands. To realize such a drastic increase in carrier mobility, the factor that dominates the scattering events should be largely modified. For example, in two-dimensional electron gas in heterostructures at low temperature, the modulation doping by inserting a spacer between carriers and impurities is crucial [59,60]. In graphene, using an atomically flat substrate successfully enhances the mobility [61]. For $\pi$-conjugated polymers, the mobility can be drastically enhanced by four orders of magnitude by an electrochemical doping, which switches the system from a hopping to a metallic transport [62]. Because scattering is mediated by the Coulomb interaction, the carrier mobility can also be sensitive to the condition of dielectric condition [63].

In the case of $Mn_3Sn$, the mean free path in equilibrium is as short as ~0.7 nm [22], which is close to the lattice constant. Considering the resistivity saturation near room temperature, the system is regarded in the Mott–Ioffe–Regel limit [22], which holds even for strongly correlated materials where the correlation localizes the carriers on each atomic site [23]. In such a system where transport is severely limited by the on-site Coulomb interaction, carrier mobility may be altered by intense photoexcitation because a large number of electrons are excited from localized to delocalized states, which screens the on-site Coulomb interaction [36–41]. For example, recently a photoinduced transition by screening the Coulomb interaction was also reported in correlated Weyl semimetal $T_d$-$MoTe_2$ using time-resolved ARPES [50], which implies that, even in a metallic system in equilibrium, the strong correlation can be effectively turned off by injecting delocalized photocarriers. Thus, a similar change in the band structure may also be expected in $Mn_3Sn$ when high-density photocarriers are injected.

The assumption that the equilibrium transport in $Mn_3Sn$ is strongly interrupted by the correlation is compatible with the ARPES results for the equilibrium state in $Mn_3Sn$ [21]. We added ARPES data for a bulk single crystal of $Mn_3Sn$ in a several-eV energy scale in Fig. 5. We present an ARPES map in a wide energy region and the corresponding energy distribution curves (EDCs) in Figs. 5(b) and 5(c). The coherent quasiparticle peak in the vicinity of $E_F$ [the red arrow in Fig. 5(c)] consists of multiple band structures and was reported in a previous study [21] to include band structures forming magnetic Weyl fermions. In addition, a much brighter incoherent part of the excitation was also clearly seen at ~3 eV away from $E_F$ (the black arrow). The latter is reminiscent of a Hubbard band of Mott-Hubbard systems where the effective mass of conduction electrons tends to be enhanced together with a decrease of spectral weight in the coherent peak with increasing correlation. In our data, it is clear that the spectral weight of the incoherent part is more dominant than that of the coherent part. Figure 5(d) shows a magnified energy-momentum view of the ARPES spectrum within the red rectangle in (b). We also point out the mass enhancement of the conduction electrons, as shown in Fig. 5(d), where the band structure calculated by DFT needs to be narrowed by a factor of 5 to explain the ARPES band dispersion in the low energy region (marked by the red rectangular in Fig. 5(b)) [21]. This band renormalization corresponds to a large mass enhancement of the conduction electrons. It is also noteworthy that trivial metallic bands, which are expected to exist at around Γ-A-L lines in the DFT calculations in Fig. 1(b), were not observed in our ARPES experiments. Thus, except for the finite quasiparticles around $E_F$, the results of APRES were very similar to that of the correlation-induced gapped systems.

Presuming that the trivial bands in $Mn_3Sn$ forms a pseudo gap by the correlation, we consider a possibility of photocarrier-induced large change in the electron band structure, similar to the nonmetal-to-metal transition, by screening the correlation [36–41]. Based on the Hubbard model, the extra carrier density required to screen the Coulomb interaction is given by a universal relation $n_{\mathrm{CD}}^{1/3} a_0 = 0.26$ [64,65], where $n_{\mathrm{CD}}$ is the extra carrier density and $a_0$ is the radius of localized electron wavefunction. This relation is known to generally hold for varieties of systems [65]. Importantly, $n_{\mathrm{CD}}$ is not determined by the actual size of Hubbard gap nor the character of wavefunctions and is only determined by the spatial size of electron localization. Assuming that the correlation in $Mn_3Sn$ forces the electrons to localize at each atomic site, $a_0$ can be regarded as a lattice constant of 5.6 Å. In this manner, $n_{\mathrm{CD}}$ is estimated to be $\sim 1\times10^{20}$ $\mathrm{cm}^{-3}$. In our experiment, the photoexcited carrier density for the threshold $I_p = 1$ mJ $\mathrm{cm}^{-2}$ was $5\times10^{20}$ $\mathrm{cm}^{-3}$, which is in reasonable agreement with $n_{\mathrm{CD}}$. Note that the photoexcited carrier density was estimated using a linear absorption coefficient. The strong pump fluence can saturate the absorption, which would further improve the agreement. Thus, the strong correlation in $Mn_3Sn$, the origin of suppressed carrier mobility in equilibrium, can be turned off by the intense photoexcitation, which may drastically alter the electronic band structure and transport properties.

The long scattering time of 300 fs is much longer than that of typical metals (~several fs), but has often been observed in topological semimetals because of suppressed backward scattering [66–69]. In addition, the density of unusual carriers in Fig. 2(f) is as low as ~$10^{17}$ cm$^{-3}$, indicating that the density of states is very small. These results imply that magnetic Weyl dispersion may be related to the unusual carriers. This interpretation seems to be consistent with the time evolution of unusual carriers and temperature. Using the two-temperature model analysis (See Appendix B), we estimated the electron and lattice temperatures for the maximum pump fluence as shown in Figs. 6(a) and 6(b). When the base temperature is 220 K, the photoexcitation elevates the system to 370 K, which remains below $T_N$. By contrast, for the base temperature of 300 K, the lattice temperature increases to 450 K within a few picoseconds after the pump, suggesting a transition into the paramagnetic phase. Thus, the magnetic octupole phase could survive for only a few picoseconds at the base temperature of 300 K. This result coincides with the experimental result in Fig. 4(c) where the unusual carriers survive in only less than a few picoseconds. Thus, the results suggest a possible relation between the unusual carriers and magnetic Weyl fermions that released from the correlation. A future study, such as direct observation of band structure using time-resolved ARPES, will prove further in-depth understanding of the emergence of unusual carriers in nonequilibrium.

In summary, we employed ultrafast time-resolved THz Faraday rotation spectroscopy to the correlated kagome bad metal $Mn_3Sn$ and found that unusual carriers with 2000-times higher mobility emerge after injecting high-density photocarriers. The long scattering time, small density, and temperature-dependent lifetime of unusual carriers imply that dielectric screening of the correlation by photocarriers unlocks the less-scattered transport under extremely nonequilibrium conditions. The results also highlight the significant role of electronic correlation on the transport of $Mn_3Sn$ in equilibrium, and provide a possible dynamical controllability of correlation by intense photoexcitation. Because the Hall conductivity spectrum is sensitive to the effective mass in the NHE as well as to the Berry curvature in the AHE, the time-resolved THz Faraday rotation spectroscopy would offer a new avenue for studying ultrafast change in the band structure to study the interplay between the non-trivial band topology and many-body interaction.

## ACKNOWLEDGEMENT

This work was supported by JST PRESTO (Grant No. JPMJPR20LA), JST FOREST (Grant No. JPMJFR2240), JSPS KAKENHI (Grants Nos. JP24K00550, JP20J01422, and JP21K13858), JST CREST (JPMJCR20R4), and JST Mirai Program (JPMJMI20A1). Near-infrared transmission and reflection spectroscopy was performed using the facilities of the Materials Design and Characterization Laboratory in the Institute for Solid State Physics, the University of Tokyo. The work at the Institute for Quantum Matter, an Energy Frontier Research Center was funded by DOE, Office of Science, Basic Energy Sciences under Award ¥# DE-SC0019331. R.M. and T.Matsuda conceived this project. T.H., H.P., T.Matsuo, and S.N. fabricated the sample and characterized it in the DC measurement. T.Matsuda performed the THz spectroscopy experiments and analysis with helps of N.K., Y.H., N.Y., J.Y., R.S., and R.M. T.Koretsune conducted the DFT calculation. K.K, C.B., and T.Kondo performed the ARPES experiment and analyzed the data. All the authors discussed the results. T.Matsuda and R.M. wrote the manuscript with substantial feedbacks from S.N., K.K. and all the coauthors.

## APPENDIX A: ANALYSIS OF HALL CONDUCTIVITY SPECTRUM

In equilibrium, the Hall conductivity in $Mn_3Sn$ is dominated by the AHE, while the NHE is negligibly small [12]. The origin of the large AHE of $Mn_3Sn$ is attributed to the intrinsic mechanism [34], which is solely determined by the electron band structure. The anomalous Hall conductivity in the dc limit is given by the integration of Berry curvature at the occupied states of electrons. In the ac regime, the anomalous Hall conductivity spectrum $\tilde{\sigma}_{\mathrm{AHE}}(\omega)$ is a complex number, and the real and imaginary parts represent nondissipative and dissipative currents, respectively. At the onset frequency of interband absorption of electrons, the imaginary part of $\tilde{\sigma}_{\mathrm{AHE}}(\omega)$ has a finite value. According to the spectral shape of the imaginary part, the real part also changes with frequency owing to the Kramers–Kronig relation. A previous THz spectroscopy has observed $\tilde{\sigma}_{\mathrm{AHE}}(\omega)$ in $Mn_3Sn$ at room temperature up to 6 THz (~25 meV) [30] and showed that the real part is almost independent of frequency and the imaginary part is negligibly small from dc to a few THz (~10 meV). This result is reasonable because the spectral shape of the interband transition is smeared by the large thermal energy of 300 K, resulting in a broadening of the onset of the imaginary part as well as a negligible frequency dependence of the real part. Recently a similar frequency-independent $\tilde{\sigma}_{\mathrm{AHE}}(\omega)$ has also been reported in other ferromagnets [31]. Thus, the contribution of AHE in Eq. (1) is given by a frequency-independent real number $\sigma_{\mathrm{AHE}}$.

Next, we consider the pump-induced change of the Hall conductivity spectrum $\tilde{\sigma}_{xy}(\omega)$ in Figs. 2(a) and 2(b). When the pump fluence is moderate (< 0.5 mJ $cm^{-2}$), a previous study reported that $\mathrm{Re}\,\tilde{\sigma}_{xy}(\omega)$ is still flat, $\mathrm{Im}\,\tilde{\sigma}_{xy}(\omega)$ is negligibly small, and only the absolute value of $\mathrm{Re}\,\tilde{\sigma}_{xy}(\omega)$ decreases by the pump. The result can be attributed to the rise of electron temperature in an ultrafast regime and the photoinduced demagnetization in slower dynamics [34]. When the pump fluence exceeds 1 mJ $cm^{-2}$, however, the pump-induced change of $\tilde{\sigma}_{xy}(\omega)$ is qualitatively different, as shown in Figs. 2(a) and 2(b); the sign flips in the real part and the peak appears in the imaginary part. In addition, the peak grows and shifts to the higher-frequency side as the external magnetic field increases. The result strongly suggests that $\tilde{\sigma}_{xy}(\omega)$ is no longer dominated by the AHE and rather described by the NHE, *i.e.,* the cyclotron resonance. The cyclotron resonance can be described by the classical Drude model coupled between the $x$- and $y$-directions:

$$m^*\ddot{x} - m^*\dot{x}/\tau_c - e\dot{y}B = -eE_x,$$
$$m^*\ddot{y} - m^*\dot{y}/\tau_c + e\dot{x}B = -eE_y.$$

By solving these equations, the normal Hall conductivity spectrum is given by

$$\tilde{\sigma}_{\mathrm{NHE}}(\omega) = \frac{\varepsilon_0\omega_p^2\omega_c}{(\omega + \mathrm{i}/\tau_c)^2 - \omega_c^2}.$$

Thus, we obtain Eq. (1) as the sum of the NHE and AHE.

Using the carrier mobility $\mu$, the normal Hall conductivity can also be written as

$$\tilde{\sigma}_{\mathrm{NHE}}(\omega) = -\frac{N_c e \mu^2 B}{(1 - \mathrm{i}\omega\tau_c)^2 + \mu^2 B^2}.$$

Thus, the NHE in the dc limit and weak-$B$ limit is proportional to $\mu^2$. This result shows the sensitiveness of NHE to the carrier mobility, which explains the clear contrast of NHE in equilibrium and nonequilibrium; In equilibrium, even though the normal carrier density is as large as $N_c$ ~ $2\times10^{22}$ $\mathrm{cm^{-3}}$ [12,22], the mobility $\mu$ is as low as 1 $\mathrm{cm^2V^{-1}s^{-1}}$. In addition, the spectral profile is blurred owing to the short scattering time of $\tau_c$ ~ 8 fs observed by THz spectroscopy [30,53], giving that $m^*$ should be as large as 11 $m_e$. Thus, $\tilde{\sigma}_{\mathrm{NHE}}(\omega)$ in $Mn_3Sn$ is hardly observed in equilibrium [53]. By contrast, the unusual carriers observed in this work have a much higher mobility $\mu$ ~ 2,600 $\mathrm{cm^2V^{-1}s^{-1}}$. Thus, even though $N_c$ is as small as ~$10^{17}$ $\mathrm{cm^{-3}}$, the unusual carriers contribute to $\tilde{\sigma}_{\mathrm{NHE}}(\omega)$ much largely than the normal carriers. In addition, the spectral feature is much sharper owing to the long scattering time $\tau_c$ ~ 300 fs. Therefore, $\tilde{\sigma}_{\mathrm{NHE}}(\omega)$ of the unusual carriers is clearly identified in this work.

## APPENDIX B: TWO-TEMPERATURE MODEL ANALYSIS

Here we consider the heating effect by the strong optical pump and its relationship with the Néel temperature using the two-temperature model. Time evolutions of the electron temperature $T_e$ and the lattice temperature $T_L$ are expressed by

$$C_e \frac{\partial}{\partial t} T_e(t) = -g\big(T_e(t) - T_L(t)\big) + S(t), \tag{B1}$$

$$C_L \frac{\partial}{\partial t} T_L(t) = g\big(T_e(t) - T_L(t)\big), \tag{B2}$$

where $C_e$=596 J $\mathrm{m^{-3}}$ $\mathrm{K^{-2}}\times T_e$ is the electron heat capacity, $C_L$=$3.3\times10^6$ J $\mathrm{m^{-3}}$ $\mathrm{K^{-1}}$ is the lattice heat capacity, and $g$=$1.8\times10^{17}$ W $\mathrm{m^{-3}}$ $\mathrm{K^{-1}}$ is the electron-phonon coupling constant. $S(t)$ is the source term given by

$$S(t) = I_p(1 - R)\frac{1}{\tau_p L_p}\exp\left\{-4\ln(2)\left(t/\tau_p\right)^2\right\}, \tag{B3}$$

where $I_p$ is the pump fluence, $R$ = 0.5 and $L_p$ = 22 nm are the reflectivity and the penetration depth for the pump photon energy of 1.55 eV, respectively. $\tau_p$ = 40 fs is the pump pulse width. Solving Eqs. (B1)–(B3), we can estimate the time-dependent electron and lattice temperature elevated by the pump. Figure 6(a) represents the time evolution of the electron and lattice temperatures at the base temperature 220 K with $I_p$ = 2.2 mJ cm$^{-2}$. At a few picoseconds after the strong pump irradiation, the lattice is heated up to 370 K, which is still below the Néel temperature. suggesting that the magnetic octupole phase can be maintained. By contrast, Fig. 6(b) shows the results at the base temperature 300 K with $I_p$ = 3.3 mJ cm$^{-2}$. the pump can heat up the lattice temperature to 450 K, which is above the Néel temperature. Thus, the magnetic octupole phase should vanish within a few picoseconds after the strong pump. This is consistent with the experimental result that the feature of unusual carriers rapidly vanishes at the base temperature of 300 K in Fig. 4(c).

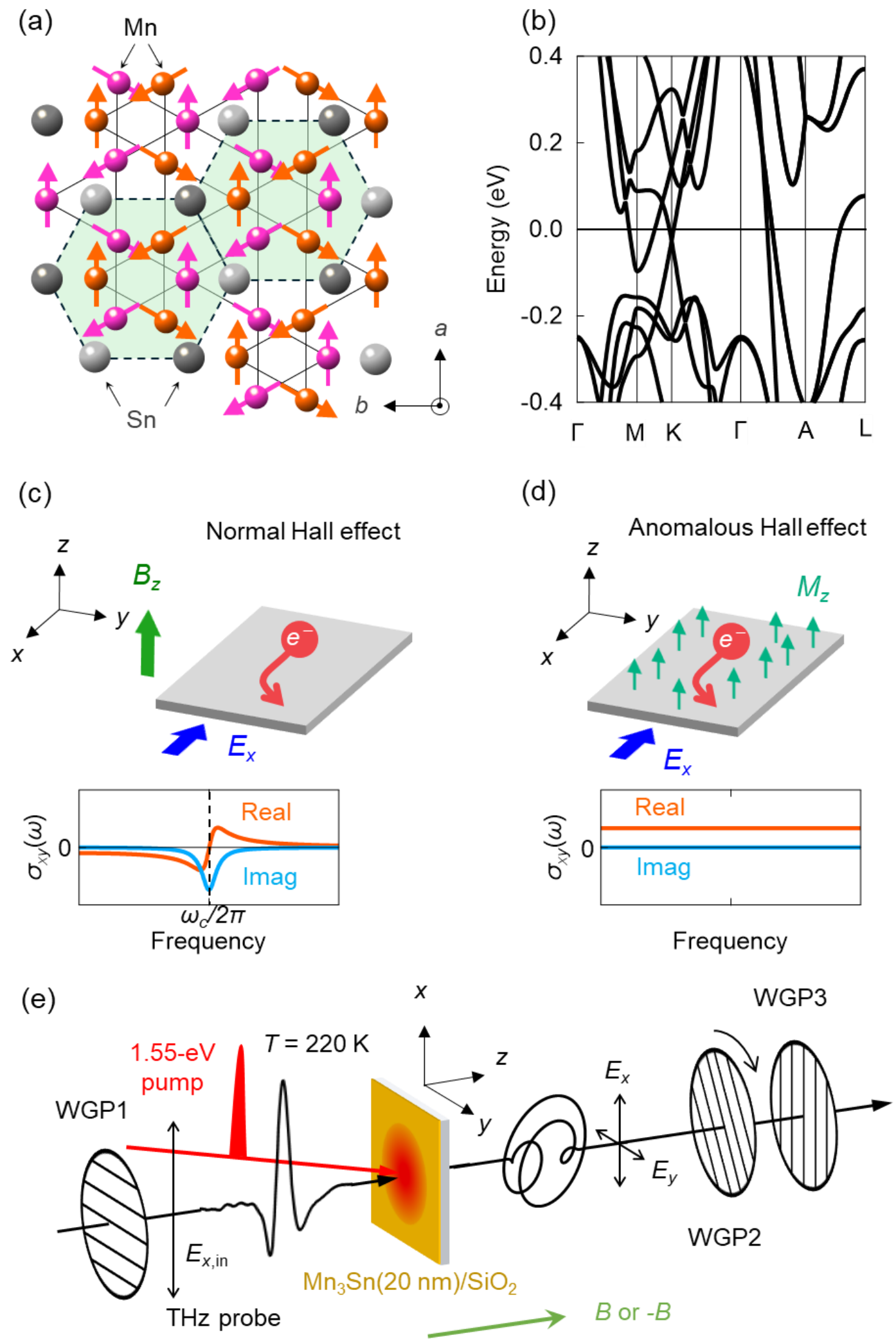

**FIG. 1.** (a) Schematic of the spin structure in $Mn_3Sn$, forming ferroic ordering of a cluster magnetic octupole (light green region). (b) DFT band structure of $Mn_3Sn$. (c)(d) Schematics of NHE and AHE, respectively. The bottom panels illustrate the Hall conductivity spectra, showing the cyclotron resonance peak in NHE (c) and non-dissipative intrinsic AHE (d). (e) Schematic of the optical pump and THz Faraday rotation probe spectroscopy. WGP: a wire-grid polarizer.

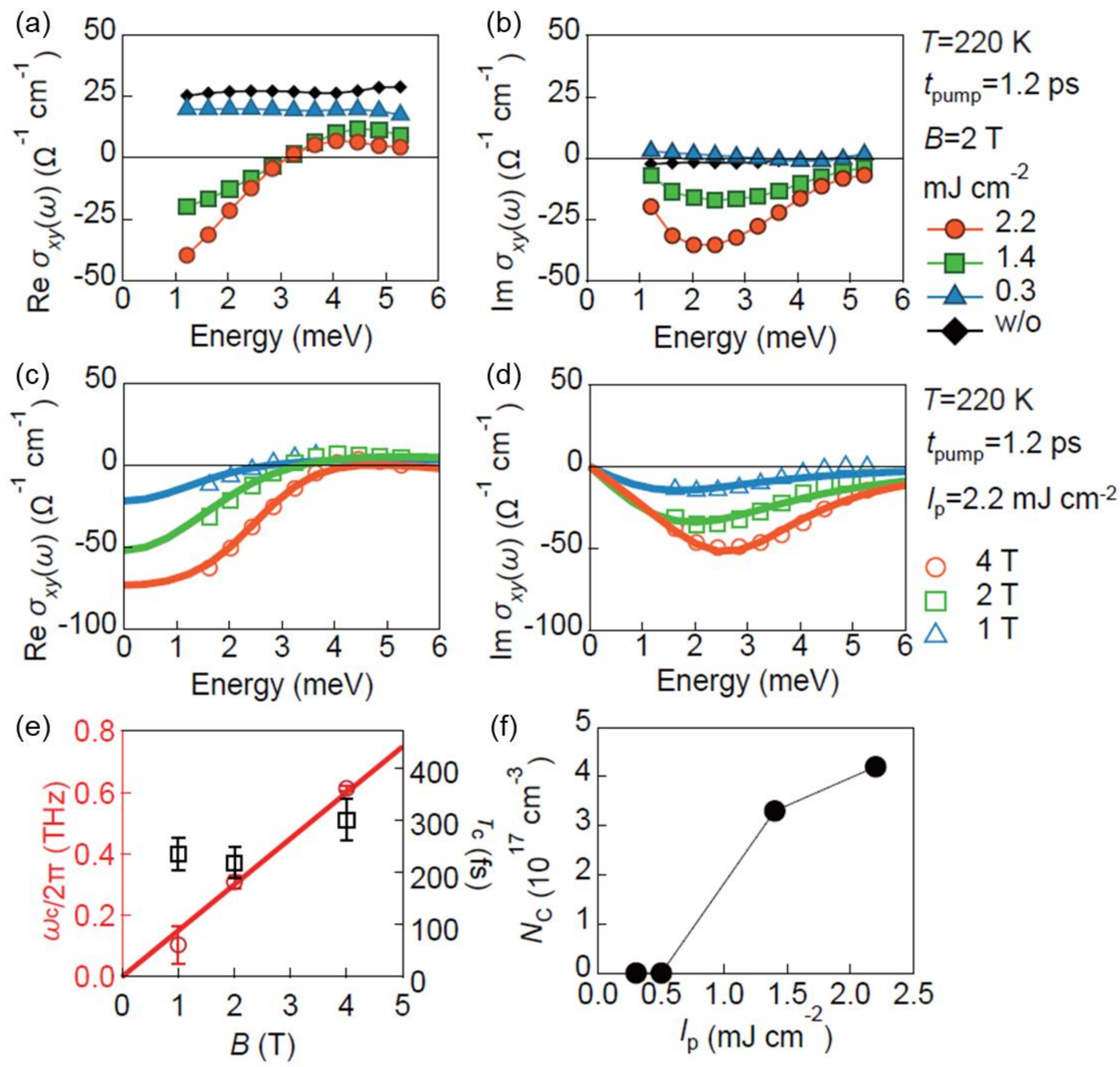

**FIG. 2.** (a)(b) Real and imaginary-part complex Hall conductivity spectra $\tilde{\sigma}_{xy}(\omega)$ at a fixed pump delay $t_{\mathrm{pump}}$ = 1.2 ps with various pump fluences $I_p$ under a magnetic field of $B$ = 2 T. (c)(d) Real and imaginary part $\tilde{\sigma}_{xy}(\omega)$ at $t_{\mathrm{pump}}$ = 1.2 ps with $I_p$ = 2.2 mJ cm$^{-2}$ under various $B$. The solid curves represent fitting data, considering the cyclotron resonance. (e) $B$ dependence of the cyclotron frequency $\omega_c/2\pi$ and the scattering time $\tau_c$. The solid line is linear fitting. (f) $I_p$ dependence of carrier density $N_c$.

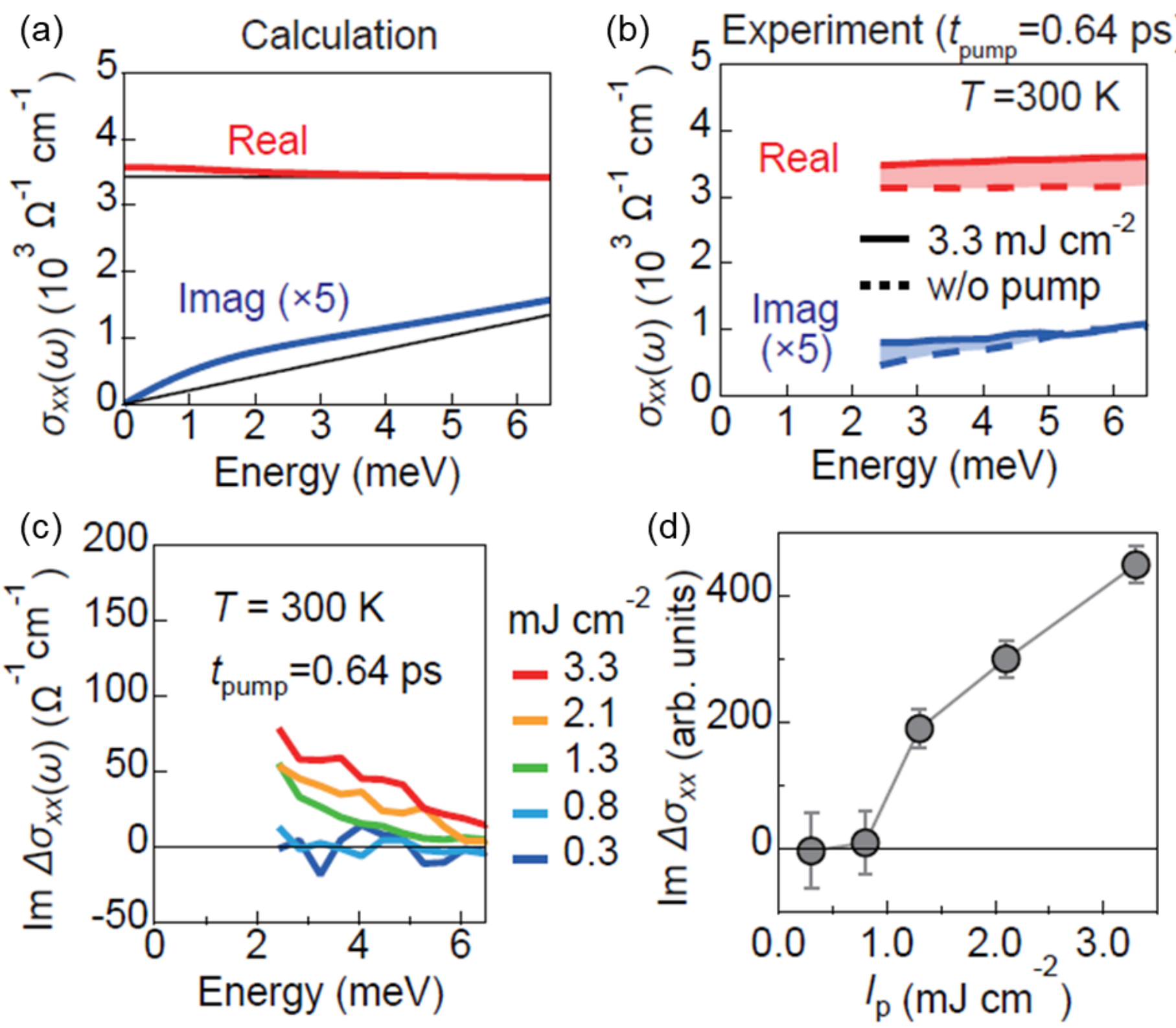

**FIG. 3.** (a) Calculated longitudinal conductivity spectra $\tilde{\sigma}_{xx}(\omega)$. The thin black curves show the contribution of original carriers (the first term in Eq. (1)). The thick red and blue curves include both the first and second terms. The imaginary parts are factored by 5 for clarity. (b) $\tilde{\sigma}_{xx}(\omega)$ at the fixed pump delay $t_{\text{pump}}$ = 0.64 ps with pump fluence $I_p$ = 3.3 mJ cm$^{-2}$ at 300 K. The dashed lines show $\tilde{\sigma}_{xx}(\omega)$ in equilibrium. (c) Pump-induced change of the imaginary part, $\text{Im}\,\Delta\tilde{\sigma}_{xx}(\omega)$, with different $I_p$. (d) $I_p$ dependence of spectrally integrated $\text{Im}\,\Delta\tilde{\sigma}_{xx}$.

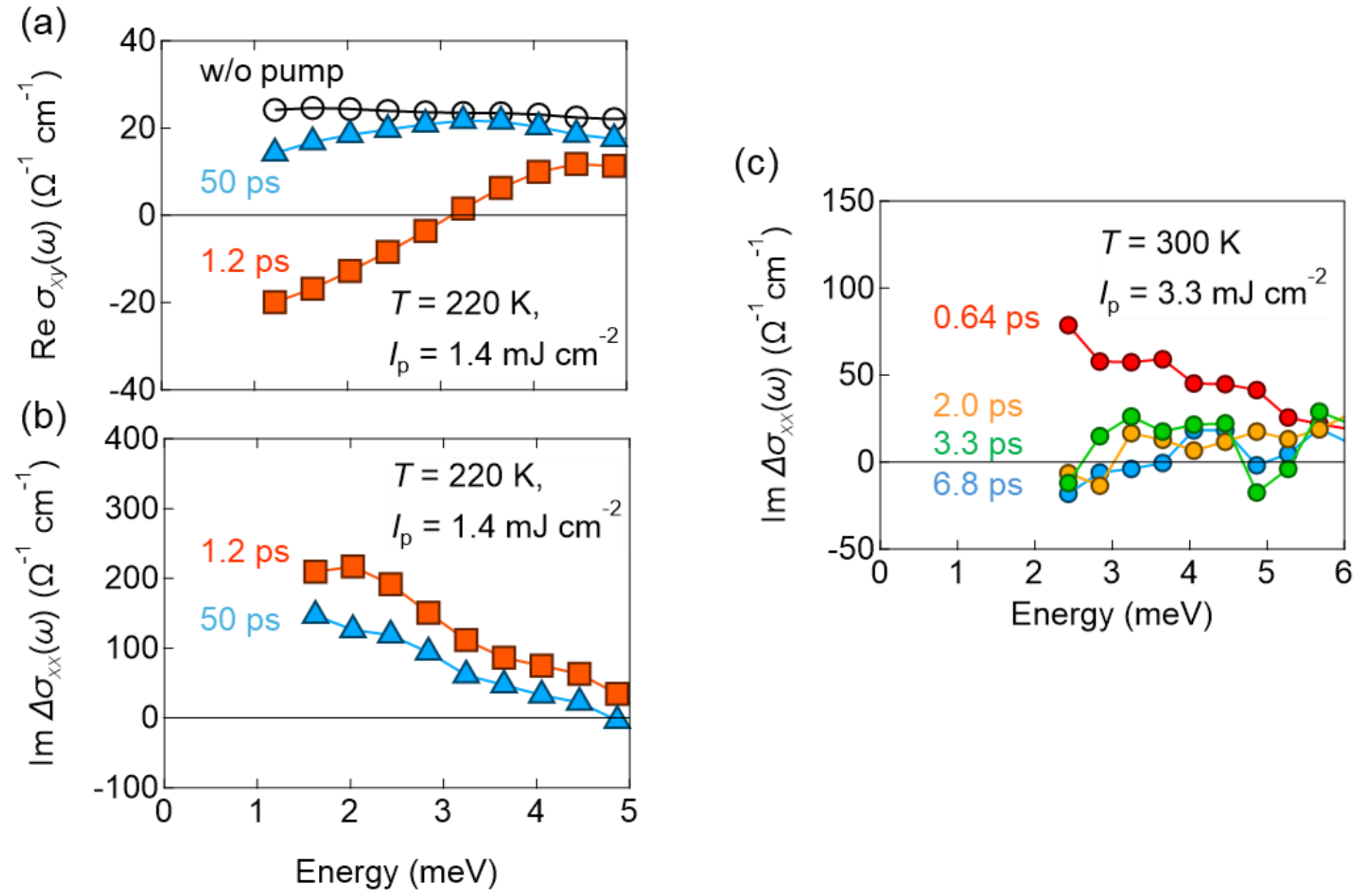

**FIG. 4.** (a) Re $\tilde{\sigma}_{xx}(\omega)$ with $I_p$ = 1.4 mJ cm$^{-2}$ at pump delays of $t_{\text{pump}}$ = 1.2 ps (orange) and 50 ps (light blue), respectively, at the base temperature of 220 K under a magnetic field of 2 T. The open circles correspond to $\tilde{\sigma}_{xy}(\omega)$ in equilibrium. (b) Im $\Delta\tilde{\sigma}_{xx}(\omega)$ with $I_p$ = 1.4 mJ cm$^{-2}$ at $t_{\text{pump}}$ = 1.2 ps (orange) and 50 ps (light blue) at the base temperature of 220 K. (c) Im $\Delta\tilde{\sigma}_{xx}(\omega)$ with $I_p$ = 3.3 mJ cm$^{-2}$ at $t_{\text{pump}}$ = 0.64, 2.0, 3.3, and 6.8 ps at the base temperature of 300 K.

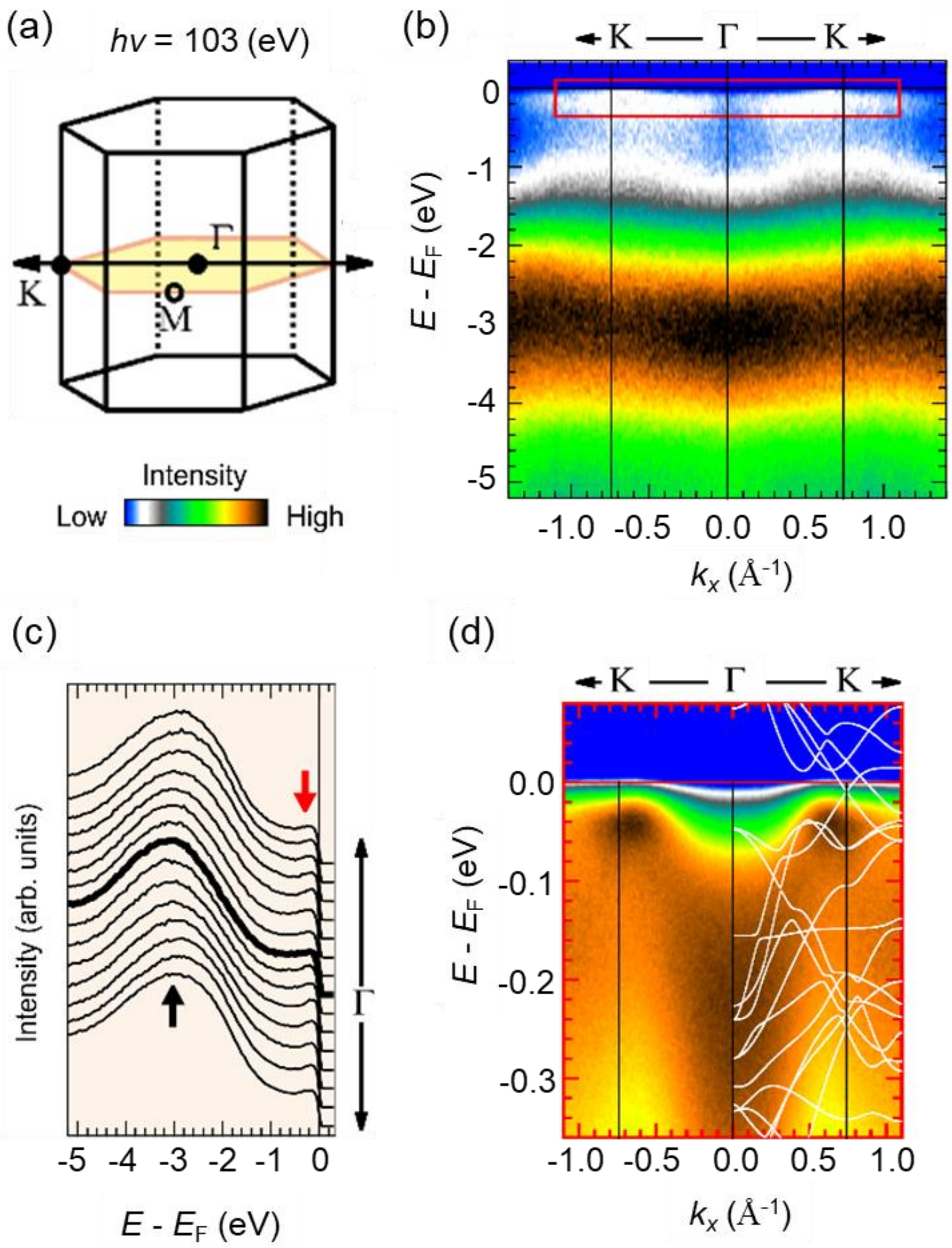

**FIG. 5.** (a) Brillouin zone with the $\boldsymbol{k}_x$-$\boldsymbol{k}_y$ sheet at $\boldsymbol{k}_z$=0 Å$^{-1}$, for which ARPES maps were taken by using a photon energy $h\nu$=103 eV. (b) ARPES map in wide energy range cut along K-Γ-K line [see an arrow in (a)]. (c) The corresponding energy distribution curves (EDCs). The bold lines indicate EDC at Γ point. Two main features are observed: coherent quasiparticle peaks at vicinity of $E_F$ (red arrow) and incoherent broad spectra around $E$–$E_F$=–3 eV (black marks). (d) The magnified energy-momentum view of the ARPES spectrum within the red rectangle in (b). The observed band features of the quasiparticle peaks are compared to the calculated bands with strong band renormalization by a factor of 5 (white lines) [21].

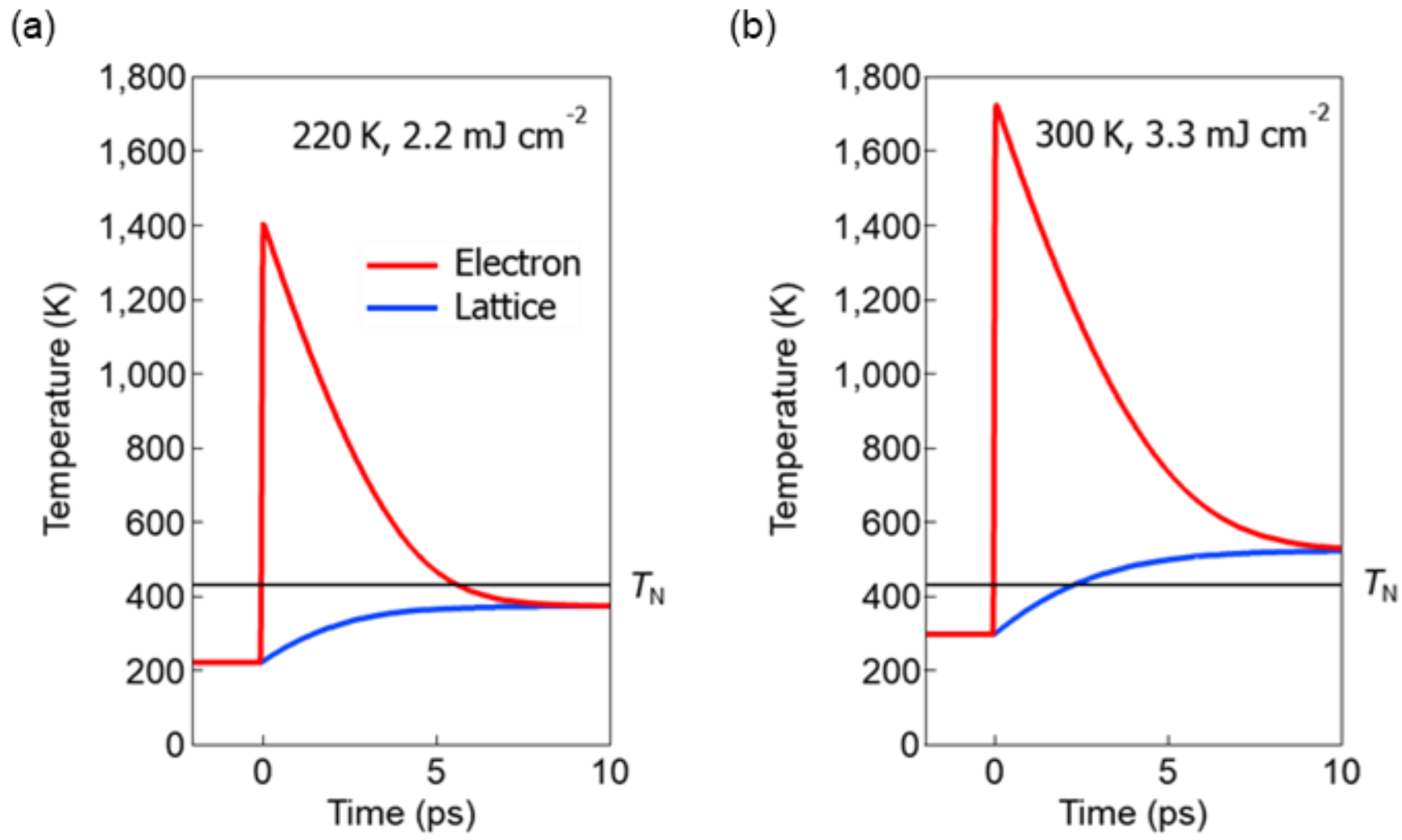

**FIG. 6** Time evolution of the electron and lattice temperatures in the cases of (a) 220 K and 2.2 mJ cm$^{-2}$ and (b) 300 K and 3.3 mJ cm$^{-2}$.